\documentclass[twocolumn]{aastex61}
\pdfoutput=1 %for arXiv submission
\usepackage{amsmath,amstext}
\usepackage[T1]{fontenc}
\usepackage[figure,figure*]{hypcap}
\usepackage{multirow}

\usepackage{color,ulem}

\newcommand{\msun}{{\rm M_{\odot}}}
\newcommand{\MKN}{{MKN}}
\newcommand{\AT}{AT2017gfo}

\newcommand{\adndt}{{\it At. Data Nucl. Data Tables}}

\shorttitle{\AT{}: an anisotropic and three-component kilonova counterpart of GW170817}
\shortauthors{Perego et al.}

\begin{document}

\title{\AT{}: an anisotropic and three-component kilonova counterpart of GW170817}

\author{Albino \surname{Perego}}
\affiliation{Istituto Nazionale di Fisica Nucleare, Sezione Milano Bicocca, gruppo collegato di Parma, Parco Area delle Scienze 7/A, I-43124 Parma, Italia}
\affiliation{Dipartimento di Fisica, Universit\`{a} degli Studi di Milano Bicocca, Piazza della Scienza 3, 20126 Milano, Italia}
\affiliation{Dipartimento di Scienze Matematiche Fisiche ed Informatiche, Universit\'a di Parma, , Parco Area delle Scienze 7/A, I-43124 Parma, Italia}
\author{David \surname{Radice}}
\affiliation{Institute for Advanced Study, 1 Einstein Drive, Princeton, NJ 08540, USA}
\affiliation{Department of Astrophysical Sciences, Princeton University, 4 Ivy Lane, Princeton, NJ 08544, USA}
\author{Sebastiano \surname{Bernuzzi}}
\affiliation{Dipartimento di Scienze Matematiche Fisiche ed Informatiche, Universit\'a di Parma, , Parco Area delle Scienze 7/A, I-43124 Parma, Italia}
\affiliation{Istituto Nazionale di Fisica Nucleare, Sezione Milano Bicocca, gruppo collegato di Parma, Parco Area delle Scienze 7/A, I-43124 Parma, Italia}

\begin{abstract}
The detection of a kilo/macronova electromagnetic counterpart (\AT{}) of the first gravitational wave signal compatible 
with the merger of two neutron stars (GW170817) has confirmed the occurrence of r-process nucleosynthesis in this kind of events. 
The blue and red components of \AT{} have been interpreted as the signature of multi-component ejecta in the merger dynamics. 
However, the explanation of \AT{} in terms of the properties of the ejecta and of the ejection mechanisms is still incomplete.
In this work, we analyse \AT{} with a new semi-analytic model of kilo/macronova inferred from general relativistic simulations 
of the merger and long-term numerical models of the merger aftermath. The model accounts for the anisotropic emission from 
the three known mass ejecta components: dynamic, winds and secular outflows from the disk. 
The early multi-band light-curves of \AT{} can only be explained by the presence of a relatively low opacity component 
of the ejecta at high latitudes. This points to the key role of weak interactions in setting the ejecta properties and 
determining the nucleosynthetic yields. Our model constrains also the total ejected mass associated to \AT{} to be between 
$0.042$ and $0.077$ $\msun$; the observation angle of the source to be between $\pi/12$ and $7\pi/36 $; and 
the mass of the disk to be $ \gtrsim 0.08 \msun$.
\end{abstract}


\section{Introduction}

%% very general intro sentence about the relevance of GW170817 and AT2017gfo
The discovery of the gravitational wave (GW) signal GW170817 by the LIGO and Virgo
collaborations has marked the beginning of the multimessenger
astronomy era \citep{Abbott2017b,GBM:2017lvda}.  
GW170817 represents not only the first observed GW signal compatible with merging binary neutron stars (BNS),
but also the first GW discovery followed by a cascade of electromagnetic (EM) signals recorded
by telescopes in space and all over the world, across the entire EM spectrum \citep{GBM:2017lvda},
from gamma-rays \citep{Abbott2017c} to radio emission \citep{Alexander2017}.

As first pointed out by \cite{Lattimer.Schramm:1974}, the ejection of
neutron star matter from a compact merger is a favorable site for the production of
the heaviest elements via the so-called r-process nucleosynthesis.
The radioactive decay of the freshly synthesized neutron-rich
r-process elements powers a transient called ``kilonova'' or ``macronova''
(\MKN{}, \citealt{Li.Paczynski:1998,Rosswog:2005,Metzger2010,Roberts.etal:2011,Kasen.etal:2013,Tanaka.Hotokezaka:2013};
the term ``macronova'' initially refered to a transient powered by free-neutron and Nickel decay, \citealt{Kulkarni:2005}).

\AT{} has been interpreted as the \MKN{} associated with GW170817. 
It peaked in less than one day after the merger in the optical and
ultraviolet bands, before fading out rapidly \citep{Nicholl2017}. 
Meanwhile, the near infrared (IR) luminosity
raised, reaching a maximum several days after the merger \citep{Chornock2017}. 
The former \MKN{} peak is called the blue component (BC), while the latter the
red component (RC). % of the \MKN{}. 
These two components could arise because of the strong dependence of the opacity 
of the material on the fraction of lanthanides and actinides. Material undergoing full 
r-process nucleosynthesis will 
produce substantial amount of lanthanides and actinides, so its presence can
explain the RC \citep{Kasen.etal:2013}. On the other hand, the BC can be explained by ejecta
that experienced only a partial r-process nucleosynthesis and is free from lanthanides and
actinides \citep{Martin.etal:2015}. The presence of a BC might be also explained by other
mechanisms, for example by the energy deposition of a relativistic jet in a
cocoon of ejecta \citep{Lazzati.etal:2017,Bromberg2017}.

%% different types of ejecta and their expected properties
For a radioactively powered \MKN,
the occurrence of a RC, of a BC or of both depends on the physical and geometrical properties of the ejecta.
Matter from a BNS merger is expected to be expelled over the whole solid angle, but not necessarily 
to be isotropic. Moreover, several ejection mechanisms play a significant role during the merger,
having a direct imprint on the ejecta properties.
%% dynamic ejecta
On the timescale of a few ms, tidal interactions and shocks drive the ejection 
of the dynamic ejecta. 
Weak interactions can potentially alter its initial neutron-richness \citep{Sekiguchi.etal:2015,Foucart.etal:2016,Radice.etal:2016,Bovard.etal:2017}.
However, equatorial ejecta seems to stay neutron-rich enough to produce
the full r-process nucleosynthesis \citep{Martin.etal:2017}.
%% wind ejecta
If the merger does not lead to the prompt collapse of the central massive neutron star (MNS),
neutrino-matter interactions and magnetic processes produce wind outflows \citep{Perego.etal:2014}. The larger timescale 
(10s~ms) and the polar character of this ejection allow matter to increase its electron fraction ($Y_{\rm e}$), 
preventing the nucleosynthesis of the heaviest r-process elements.
%% secular ejecta
If during the merger a disk has formed, on its longer lifetime
(100s~ms) the spreading due to viscous processes and the subsequent nuclear recombination
unbind a fraction of the disk \citep{Metzger2009,Fernandez.Metzger:2013,Metzger.Fernandez:2014,Just.etal:2015a}. 
Numerical studies of this secular ejecta
revealed rather homogeneous distributions of the ejecta properties, emitted at all latitudes 
and leading to full r-process nucleosynthesis \citep{Wu.etal:2016,Siegel.Metzger:2017}.

Recent work on the interpretation of \AT{} has revealed that a single
component model for the \MKN{} is inadequate to reproduce the observed
features in all bands. Two or even three component models are
necessary \citep{Cowperthwaite2017,Tanvir.etal:2017,Tanaka2017}.
In all of them a fast, low opacity ejection is responsible for the
BC, while a slower, more opaque ejection accounts for the RC
\citep{Abbott2017}.
The most sophisticated \MKN{} models are based on radiative transport schemes, 
nonetheless most of them assume a spherical geometry.

In this work, we show that the BC and the RC of the
\MKN{} in all relevant bands can be explained by an anisotropic model with
multi-component ejecta.
Our model builds on general-relativistic merger simulations 
and aftermath simulations of neutrino and viscosity-driven ejecta,
and it directly relates the geometry and the physical ejection mechanisms 
to the light curves.
By reproducing the observed light curves we confirm that the ejecta
properties are highly anisotropic and inhomogeneous. Moreover, we
constrain some of the properties of the merging system  
and prove the central role of weak interaction in BNS mergers.

\section{Anisotropic three-component semi-analytical \MKN{} model}

We propose a semi-analytical \MKN{} model with dependence on the polar
angle and composed of three ejecta
components: the dynamic, the wind and the secular ejecta. 
The ejecta propagation and the electromagnetic radiation are computed
by an extension of the model presented in \citet{Grossman.etal:2014} 
and \citet{Martin.etal:2015}.

\subsection{Ejecta components}

\begin{figure*}
  \label{fig-nr}
  \centering
  \includegraphics[width = 0.49 \linewidth]{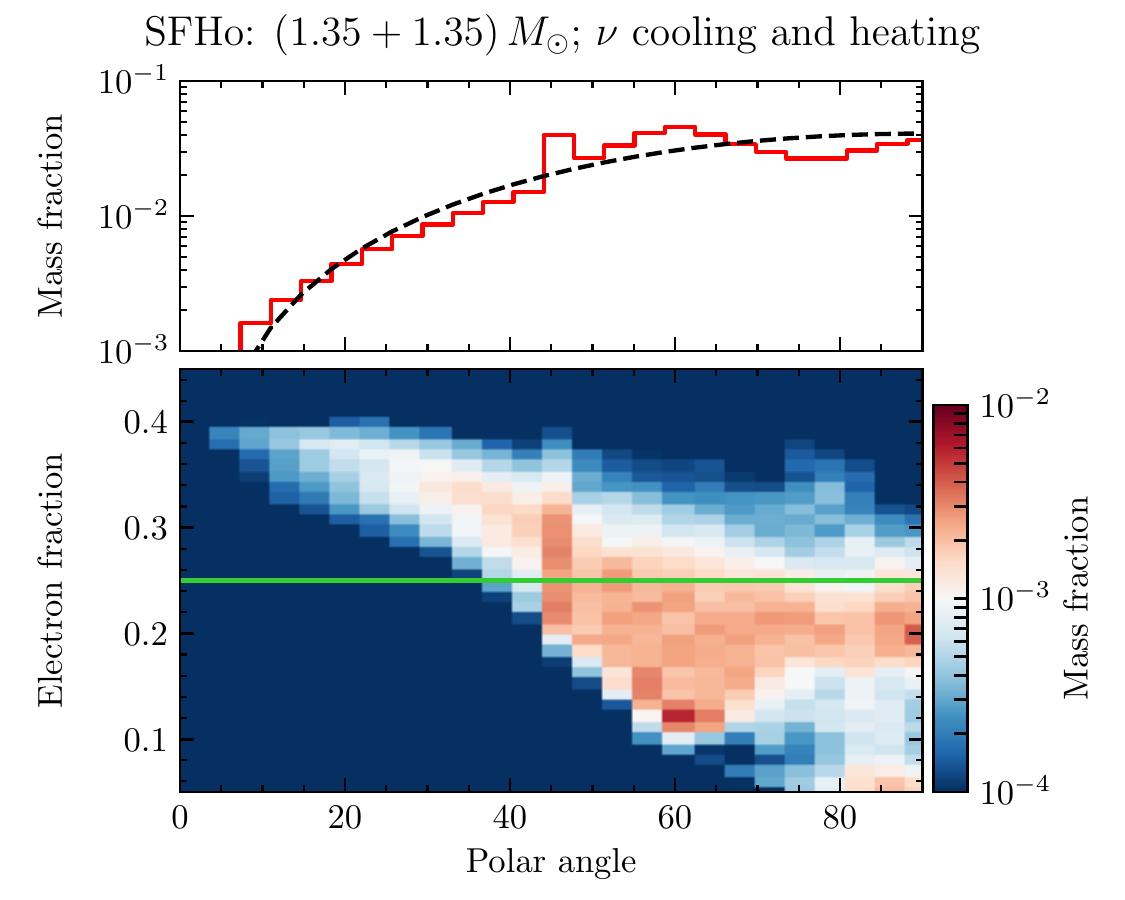}
  \includegraphics[width = 0.49 \linewidth]{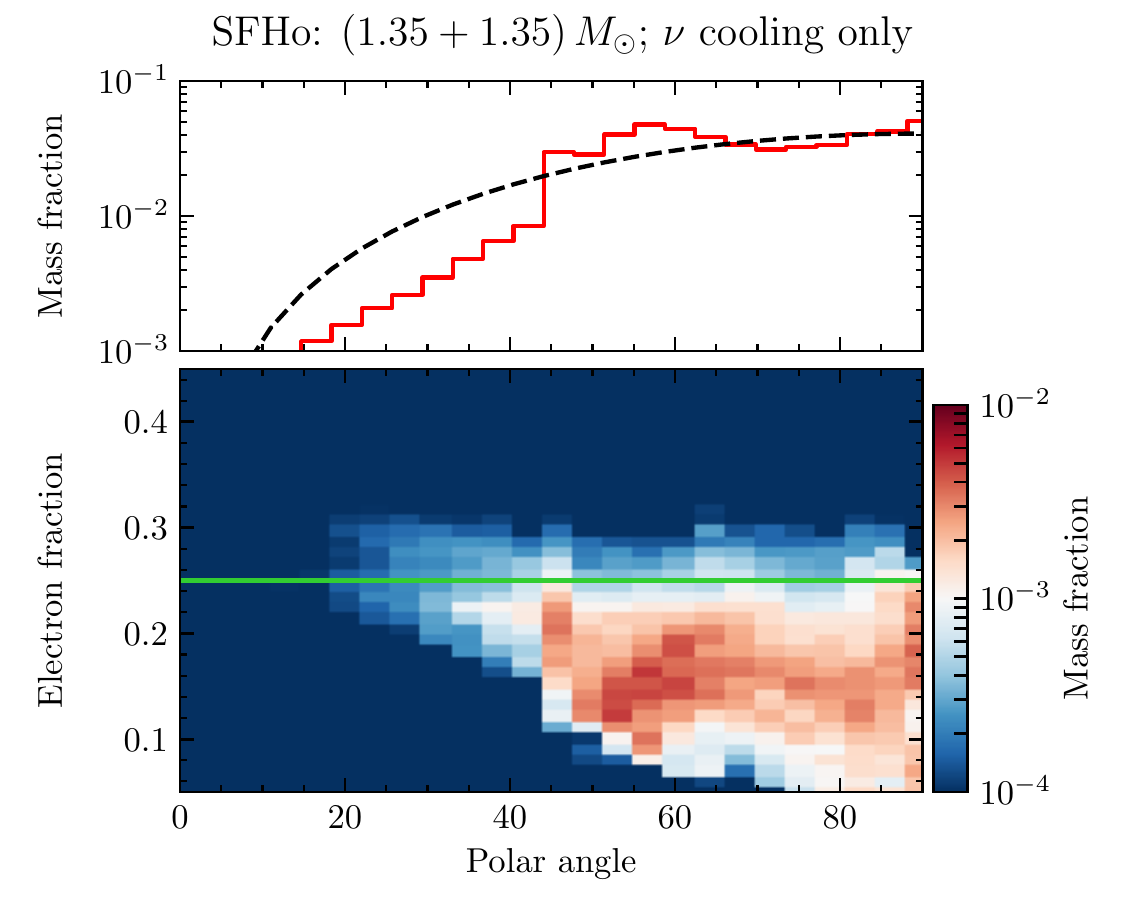}
  \caption{Angular profile and composition of dynamic ejecta
    computed from GRHD BNS merger
    simulations of an equal-mass binary of $2.7\msun$
    with SFHo EOS (Radice et al.~in prep.~2017).
    Left: Simulation with neutrino cooling and heating. 
    Right: Simulation with neutrino cooling only.
    The small bump in the ejecta distribution at $45^\circ$ is a
    an artifact imprinted by our Cartesian grid, which preferentially
    channels flow along its symmetry directions \citep{Radice.etal:2016}.
    The angular profile of the simulation including neutrino cooling and heating 
    is well described by $\sin^2\theta$.} 
\end{figure*}

\begin{figure}
 \label{fig0}
 \centering
 \includegraphics[width = 0.9 \linewidth]{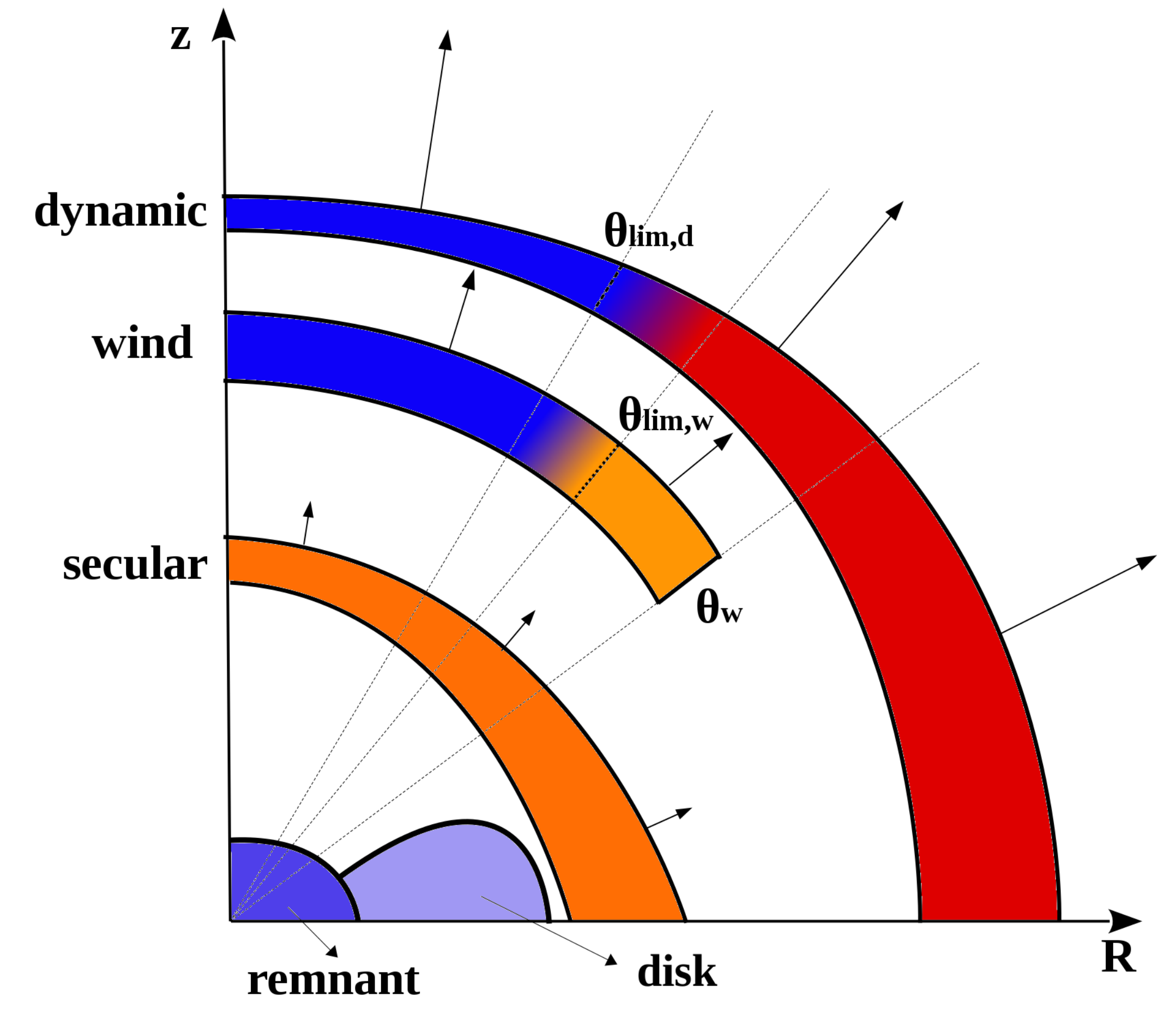}
 \caption{Graphical sketch of the three ejecta components radially expanding from the remnant.
 Different colors correspond to different matter opacity:
 high (red), intermediate (orange), low (blue).}
\end{figure}

\paragraph{Dynamic.} 
The geometry and composition of the dynamic ejecta is
inferred by general-relativistic hydrodynamical (GRHD) simulations of
Radice et al.~in prep.~(2017).
Figure~\ref{fig-nr} displays the angular distributions of the mass and $Y_{\rm e}$ for an exemplary case. 
The presence of shocks and intense neutrino irradiation increases
$Y_{\rm e}$ above 0.25 close to the polar axis while the electron fraction 
stays below $0.25$ across the equatorial plane.
The crucial impact of neutrino absorption is visible by comparing
simulations with and without neutrino heating.
The ejection happens at all latitudes, but predominantly 
along the equatorial plane. Once neutrino heating is taken into account,
the angular distribution is well approximated by $F(\theta) =
\sin^2{\theta}$.
We remark that, according to our simulations, the overall geometry
and composition of the outflow are insensitive to the nuclear equation
of state (EOS) and the binary parameters, at least up to the mass ratios
we have considered ($q \gtrsim 0.85$). On the other hand, the total
dynamic ejecta mass ranges from ${\sim}10^{-4}\msun$ to
${\sim}10^{-2}\msun$. For this reason, only the formers are used to
inform our \MKN{} model.
The ejecta velocity is mildly
relativistic, $v_{\rm d} \lesssim 0.3 c$. If the merger does not lead to the prompt
collapse of the massive neutron star (MNS) to a black hole (BH), tidal torques and mass ejection episodes from 
the rotating MNS produce a disk, with a mass $10^{-2} \msun \lesssim M_{\rm disk} \lesssim 10^{-1} \msun$.

\paragraph{Wind.}
The wind geometry is inferred from merger aftermath simulations of 
\citet{Martin.etal:2015}. The ejection mechanisms favor
polar emission with a rather uniform distribution in mass ($F(\theta) \approx {\rm const}$ 
for $\theta \lesssim \theta_{\rm w} \approx \pi/3$)
and velocity $v_{\rm w} \lesssim 0.08 c$. The ejected mass is a fraction of the disk mass, $m_{\rm ej,w} = \xi_{\rm w} M_{\rm disk}$
with $\xi_w \sim 0.05$. Neutrino irradiation has enough time to unbind matter from the disk and to increase $Y_{\rm e}$ above 0.25, 
preventing full r-process nucleosynthesis.

\paragraph{Secular.}
The properties of secular ejecta are inspired by simulations of disks 
around a MNS or a BH \citep{Wu.etal:2016,Lippuner.etal:2017,Siegel.Metzger:2017}.
This ejecta is expected to unbind a significant fraction of the disk
$m_{\rm ej,v} = \xi_{\rm v} M_{\rm disk}$, where $\xi_w \lesssim 0.3$, with a rather uniform velocity distribution and 
$v_{\rm s} \lesssim 0.05 \, c$. We consider an equatorial-dominated flow, $F_{\rm s}(\theta)=\sin^2{\theta}$ and two cases
for the $Y_{\rm e}$ distribution: $ 0.1 \lesssim Y_{\rm e}(\theta)
\lesssim 0.4$, for a MNS collapsing to a BH on a timescale shorter
than the disk lifetime \citep{Siegel.Metzger:2017}, and $ 0.25 \lesssim
Y_{\rm e}(\theta) \lesssim 0.5$ for an extremely long-lived MNS \citep{Lippuner.etal:2017}.

\begin{table*}%[h,t]
\begin{center}
  \caption{Left: Parameters for the exploration of the model. Right: Parameters of the best fits to
    \AT{}. \label{tab: best fit parameters}}
  \begin{tabular}{ll|lll}
    \tableline \tableline
               {~}                              & Parameter range & BF        &  BF$_{\rm c}$ &  BF$_{\rm c,\epsilon}$                 \\ \tableline
               $\chi^2$                         & - & 759     & 1263         & 1448                   \\
               $M_{\rm disk}$ [$\msun$]         & $\left\{ 0.01; \, 0.08; \, 0.1; \, 0.12; \, 0.15; \, 0.2  \right\} $ & 0.08      & 0.1          & 0.12                      \\
               $m_{\rm ej,d}$ [$10^{-2}\msun$]  & $\left\{ 0.05; \, 0.5; \, 1.0;\, 2.0;\, 5.0  \right\} $ & 1.0       & 0.5           & 0.5      \\
               $\xi_{\rm w}$                    & $\left\{ 0.001;\, 0.05;\, 0.1;\, 0.15;\, 0.2  \right\} $ & 0.001     & 0.15          & 0.2                       \\ 
               $\xi_{\rm s}$                    & $\left\{ 0.001;\, 0.1;\, 0.2;\, 0.3;\, 0.4  \right\} $ & 0.4       & 0.2           & 0.4                      \\ 
               $\theta_{\rm lim,d} $            & $\left\{ \pi/6;\, \pi/4 \right\}$ & $\pi/4$   & $\pi/6$       & $\pi/6$                           \\
               $\theta_{\rm lim,w} $            & $\left\{ \pi/6;\, \pi/4 \right\}$ & $\pi/6$   & $\pi/6$       & $\pi/4$   \\               
               $v_{\rm rms,d} \, [c]$           & $ \left\{ 0.1; \,0.13;\, 0.17;\, 0.2;\, 0.23 \right\} $ & 0.2       & 0.23           & 0.2   \\
               $v_{\rm rms,w} \, [c]$           & $ \{ 0.033; \,0.05;\, 0.067 \}$ & 0.067     & 0.067         & 0.067                          \\
               $v_{\rm rms,s} \, [c]$           & $ \{ 0.017;\, 0.027; \,0.033;\, 0.04 \}$  & 0.027     & 0.04          & 0.04                 \\               
               $\kappa_{\rm d}$ [${\rm cm\,g^{-1}}$] & $\{ (0.5,30);\,(1,30) \}$ & (1,30)  & (1,30)        & (1,30) \\  
               $\kappa_{\rm w}$ [${\rm cm\,g^{-1}}$] & $ \{ (0.5,5); \, (0.1,1) \} $ & (0.1,1) & (0.5,5)     & (0.5,5)                         \\
               $\kappa_{\rm s}$ [${\rm cm\,g^{-1}}$] & $\{ 1; \, 5; \, 10; \, 30 \}$ & 1       & 5           & 5 \\               
               $\theta_{\rm obs} $              & $n \, \pi/36$ for $ \, n=0 \, \dots \, 11$ & $\pi$/12  & $5 \pi/36$       & $7 \pi/36$                   \\
               $\epsilon_{o}$[$10^{18}{\rm erg \, g^{-1} \, s^{-1}}$] & $\left\{ 2; \, 6; \, 12; \, 16; \, 20 \right\}$ & 16 & 20 & 12               \\
               \tableline
\tableline
\end{tabular}
\tablecomments{BF: best fit parameter set. BF$_{\rm c}$: best fit parameter set once 
$M_{\rm disk} \leq 0.12~\msun$ , $m_{\rm ej,d} \leq 0.01~\msun$ and $\kappa_{\rm s} \geq 5.0~{\rm cm^2 g^{-1}}$ are imposed. BF$_{\rm c,\epsilon}$:
best fit when $\epsilon_0 \leq 12 \times 10^{18}{\rm erg \, g^{-1} \, s^{-1}}$ is also imposed.}
\end{center}
\end{table*}

\subsection{Ejecta expansion and radiative model}

We assume the ejecta to be axisymmetric around the rotational axis of the remnant
and symmetric with respect to the equatorial plane.
The polar angle $\theta$ is discretized in 12, equally spaced bins.
Each mass ejection is characterized by a) its mass, $m_{\rm ej}$ , b) its rms radial speed, $v_{\rm rms}$,  
c) its opacity, $\kappa$; alongside with their angular distributions. For the mass, we introduce a distribution $F(\theta)$ such that:
\begin{equation}
 m_{\rm ej} = \sum_{k=1,12} \, m_{\rm ej,k} = 
 \sum_{k=1,12} 2 \pi \int_{\theta_k - \Delta \theta/2}^{\theta_k + \Delta \theta/2} \, F(\theta) \sin{\theta} \, {\rm d}\theta \, .
\end{equation}
For $v_{\rm rms}$, we assume $v_{\rm rms}(\theta)={\rm const}$.
We assign the opacity according to the value of $Y_{\rm e}$ for the bulk of the ejecta.
If $Y_{\rm e}$ is such that $Y_{\rm e}(\theta) \gtrsim 0.25$ for
$\theta < \theta_{\rm lim}$ and $Y_{\rm e}(\theta) \lesssim 0.25$ for
$\theta > \theta_{\rm lim}$, then we set 
$\kappa(\theta > \theta_{\rm lim}) = \kappa_{\rm max} \gtrsim 10 \, {\rm cm^2 \, g^{-1}}$ 
and $\kappa(\theta < \theta_{\rm lim}) = \kappa_{\rm min} \lesssim 1 \, {\rm cm^2 \, g^{-1}}$.
Otherwise, if for all $\theta$ angles $Y_{\rm e}$ has a broad distribution across 0.25, we assign $\kappa(\theta) = \kappa_{\rm avg}$ 
with $\kappa_{\rm min} \lesssim \kappa_{\rm avg} \lesssim \kappa_{\rm max}$.

Within each bin, we run the radial model of \citet{Grossman.etal:2014} for each ejecta component.
We further assume that the energy emitted by the two innermost photospheres 
is deposited at the basis of the outermost shell or inside its radiating envelope,
and gets quickly reprocessed and emitted by the outermost photosphere.
The energy that powers the \MKN{} is expressed as $Q = \Delta M_{\rm env} \, \epsilon_{\rm nuc}$
where $\Delta M_{\rm env}$ is the mass of the radiating shell enclosed between the diffusion and the free streaming photosphere, $R_{\rm ph}$.
The nuclear heating rate, $\epsilon_{\rm nuc}$, is approximated by an analytic fitting formula, derived from detailed nucleosynthesis
calculations \citep{Korobkin.etal:2012}:
\begin{equation}
  \epsilon_{\rm nuc}(t) = \epsilon_0 \, \epsilon_{Y_{\rm e}}(t) \, \left( \frac{\epsilon_{\rm th}}{0.5} \right) \left[ \frac{1}{2} - 
    \frac{1}{\pi}\arctan{\left( \frac{t-t_0}{\sigma} \right)}   \right] \, ,
\end{equation}
where $\sigma = 0.11 \, {\rm s}$, $t_0 = 1.3 \, {\rm s}$, and $\epsilon_{\rm th}$ is the thermalization efficiency 
(Table~1 and Equation~(36) of \cite{Barnes.etal:2016}).
\cite{Korobkin.etal:2012} found $\epsilon_0 = 1.2 \times 10^{18}{\rm erg \, g^{-1} \, s^{-1} }$ using the finite
range droplet model (FRDM, \citealt{Moeller.etal:1995}). Due to the large uncertainties in the nuclear mass 
and decay models, we consider $\epsilon_0$ as a free parameter with 
$ 2 \times 10^{18} {\rm erg \, s^{-1} \, g^{-1}} \lesssim \epsilon_0 \lesssim 2 \times 10^{19} {\rm erg \, g^{-1} \, s^{-1}} $
(e.g. \citealt{Mendoza-Temis.etal:2015,Rosswog.etal:2017}).
Detailed calculations of neutrino-driven wind nucleosynthesis revealed the 
dominant presence of first r-process peak nuclei with a decay half-life of a few hours
(Table 1, \citealt{Martin.etal:2015}). The associated specific heating rate 
showed that $\epsilon_{\rm nuc}$ can significantly differ from the one of extremely neutron rich ejecta 
for $Y_{\rm e} \gtrsim 0.25$ (Figure 13, \citealt{Martin.etal:2015}). 
Thus, we have introduced the factor $\epsilon_{Y_{\rm e}}$ such that
$\epsilon_{Y_{\rm e}}(t) = \epsilon_{\rm min} + \epsilon_{\rm max} \, \left\{ 1 + \exp{\left[ 4(t/t_{\epsilon} -1) \right] }\right\}^{-1}$, with 
$t_{\epsilon} = 1 \,{\rm d}$, $\epsilon_{\rm min}=0.5$ and $ \epsilon_{\rm max} = 2.5$ if $Y_{\rm e} \gtrsim 0.25$, 
and $\epsilon_{Y_{\rm e}}(t) = 1$ otherwise.

We locate an observer at a distance $d \gg R_{\rm ph}$, with a viewing angle $\theta_{\rm obs}$
measured from the symmetry axis. From the analysis of GW170817, we set
$d=40 \,{\rm Mpc}$ and $\theta_{\rm obs} < \pi \, 11/36$ \citep{Abbott2017b}.
The observed spectral flux is computed as a superposition of Planckian distributions 
weighted by the projection of the emitting surface along the view line 
(Equations 4 and 5 in \citealt{Martin.etal:2015}).

Figure~\ref{fig0} provides a 
sketch of the system geometry and of the opacity properties of the ejecta.
The three-component anisotropic matter ejection produces a rich light curve
with peaks proceeding from the ultraviolet to the near-IR on a timescale of a few days.
The presence of a low opacity ejecta close to the polar axis produces a BC
with a peak within the first day after the merger.
The RC of the emission is produced by all ejecta, with
a dominant contribution from the more opaque dynamic ejecta and the
more abundant secular ejecta. The wind contributes to both components, 
substaining the BC for a timescale $\sim 1~{\rm d}$ and, at the same time, powering 
the first phase of the RC.

\section{Results}

\begin{figure*}
 \label{fig3}
 \begin{center}
 \includegraphics[width=0.45 \linewidth]{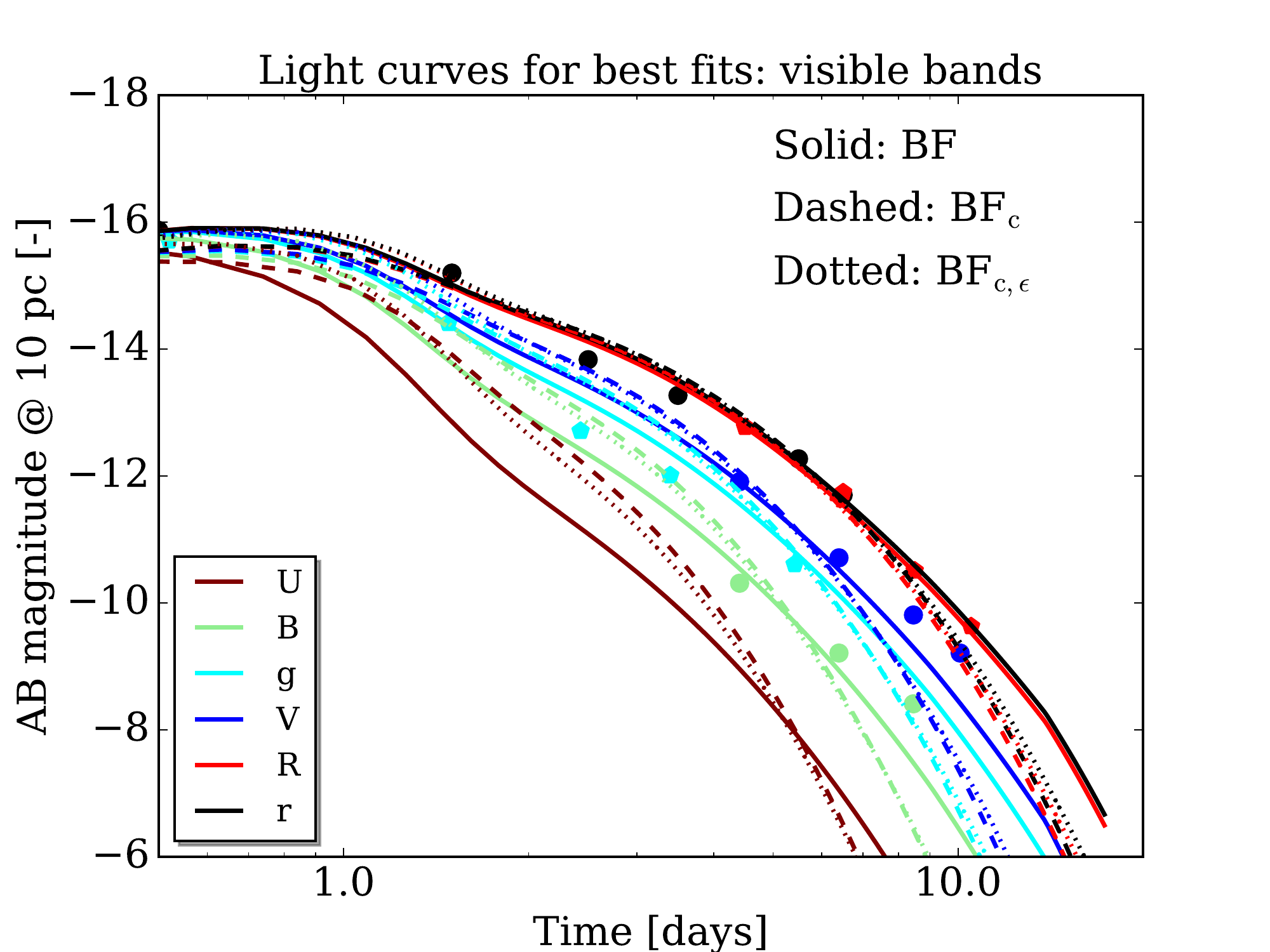}
 \includegraphics[width=0.45 \linewidth]{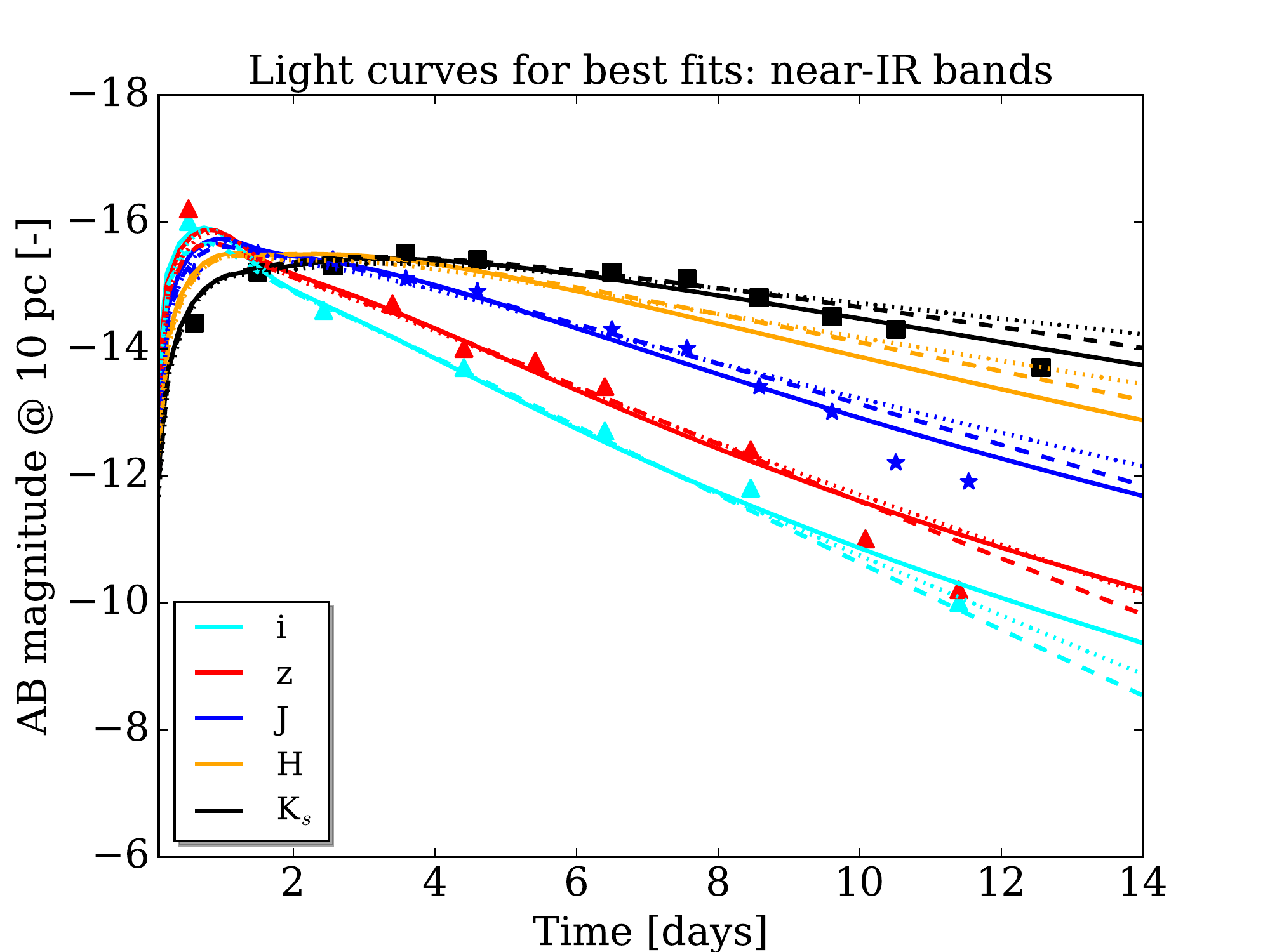}
 \end{center}
 \caption{Visible (left) and near-IR (right) light curves obtained for the best fit models reported in Table~\ref{tab: best fit parameters}:
 BF (solid), BF$_{\rm c}$ (dashed), BF$_{\rm c,\epsilon}$ (dotted).}
\end{figure*}

We compare the light curves of \AT{} in the visible and
near-IR with predictions of our model. The observed apparent
magnitudes are taken from \citet{Pian.etal:2017} for the photometric
filters $B$, $V$, $R$, $r$, $g$, $i$, and $z$; and from
\citet{Tanvir.etal:2017} for the filters $K_s$ and $J$.  

The parameter space of our model is explored by constructing a
discrete grid for each parameter as indicated in Table~\ref{tab: best fit parameters}.
The agreement between a model (specified by a set of parameters) and the data is quantified using the function,
% \begin{equation}
%   \chi^2 = \sum_{n=1}^{N_{\rm fts}} \chi^2(n) = 
% \sum_{n=1}^{N_{\rm fts}} \frac{1}{N_{{\rm pts}}(n)} \left(  \sum_{k=1}^{N_{{\rm pts}}(n)} \left( m_{k,n}^{\rm obs} - m_{k,n}^{\rm mod}   \right)^2 \right) \, ,
%  \label{eq: chi2 eq}
% \end{equation}
\begin{equation}
  \chi^2 = \sum_{n=1}^{N_{\rm fts}}   
  \left(  
  \sum_{k=1}^{N_{{\rm pts}}(n)} 
  \left( 
  \frac{m_{k,n}^{\rm obs} - m_{k,n}^{\rm mod}}{\sigma_{k,n}^{\rm obs}}   
  \right)^2
  \right) \, ,
 \label{eq: chi2 eq}
\end{equation}
where $N_{\rm fts}=9$ is the number of filters used in the comparison, $N_{{\rm pts}}(n)$ the number of points in each light curve, 
$m_{k,n}^{\rm mod}$ the apparent magnitudes given by our model and $m_{k,n}^{\rm obs} \pm \sigma_{k,n}^{\rm obs}$ the observed
apparent magnitudes with their uncertainties.
Three physically motivated best fit models are discussed in the following.

We first assume no constraints on the parameters within our grid. 
Among all the models, we found a minimum for $\chi^2 = 759$.
% , corresponding to an average discrepancy of
% $\chi^2/N_{\rm fts} \approx 0.077$ between the model and the observed magnitudes. 
% This value is of the order of the uncertainties on the observations.
The corresponding parameter set is reported as BF in Table~\ref{tab: best fit parameters} and
the light curves are represented in Figure~\ref{fig3} (solid lines).
This model is characterized by a subdominant wind component
and by a secular ejecta whose opacity is such that $\kappa_{\rm s} \ll \kappa_{\rm d,max}$, 
but $\kappa_{\rm s} \sim \kappa_{\rm w,max}$.
Thus, our three-component model has reduced to an effective two-component model 
in which the formation of a significant fraction of the heaviest r-process elements 
is not expected inside the secular ejecta. The lower $\kappa_{\rm s}$
is compensated by a slower expansion, to reproduce the features of the RC.

\begin{figure*}
 \label{fig1}
 \centering
 \includegraphics[width=0.32 \linewidth]{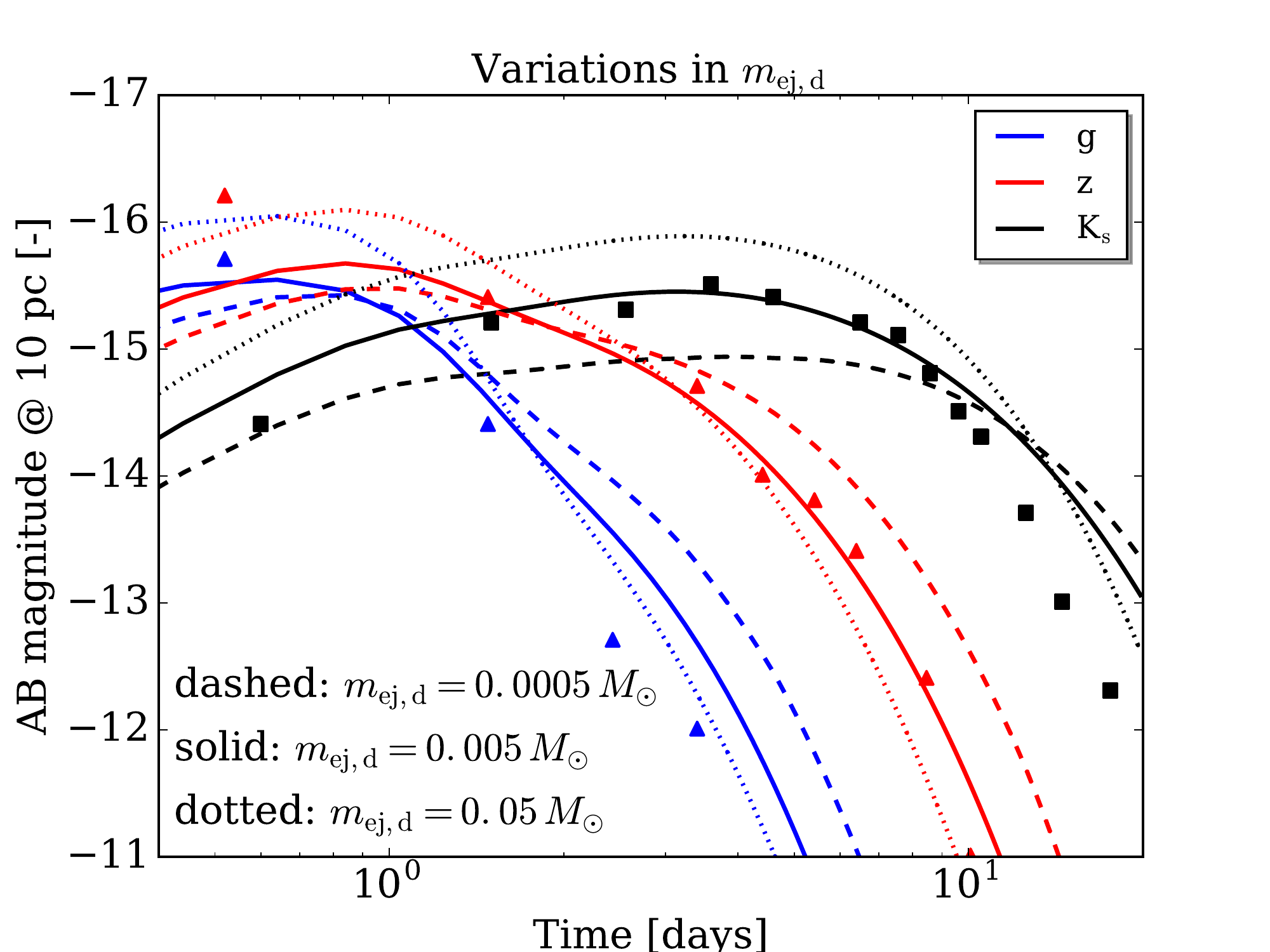}
 \includegraphics[width=0.32 \linewidth]{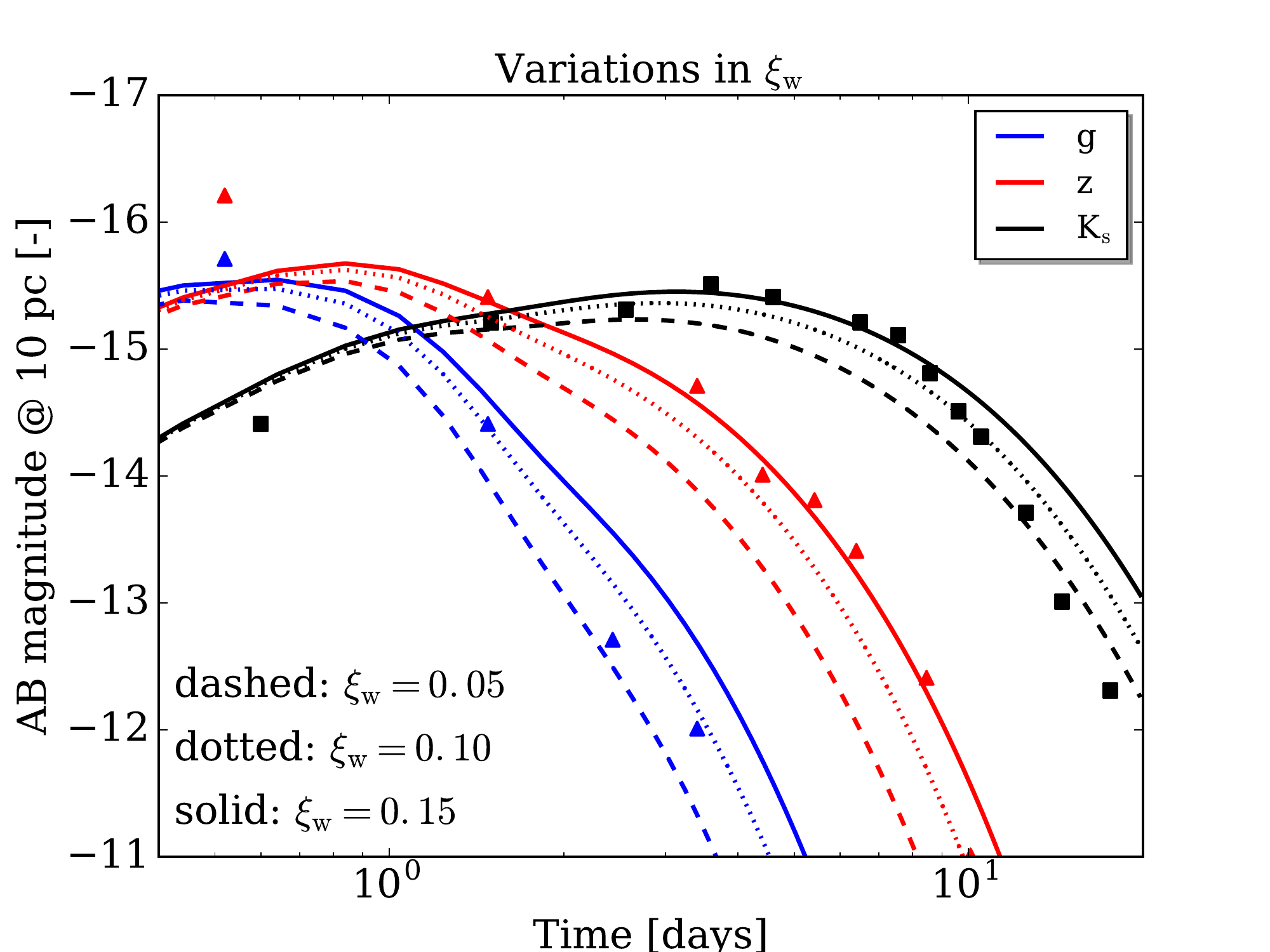} 
 \includegraphics[width=0.32 \linewidth]{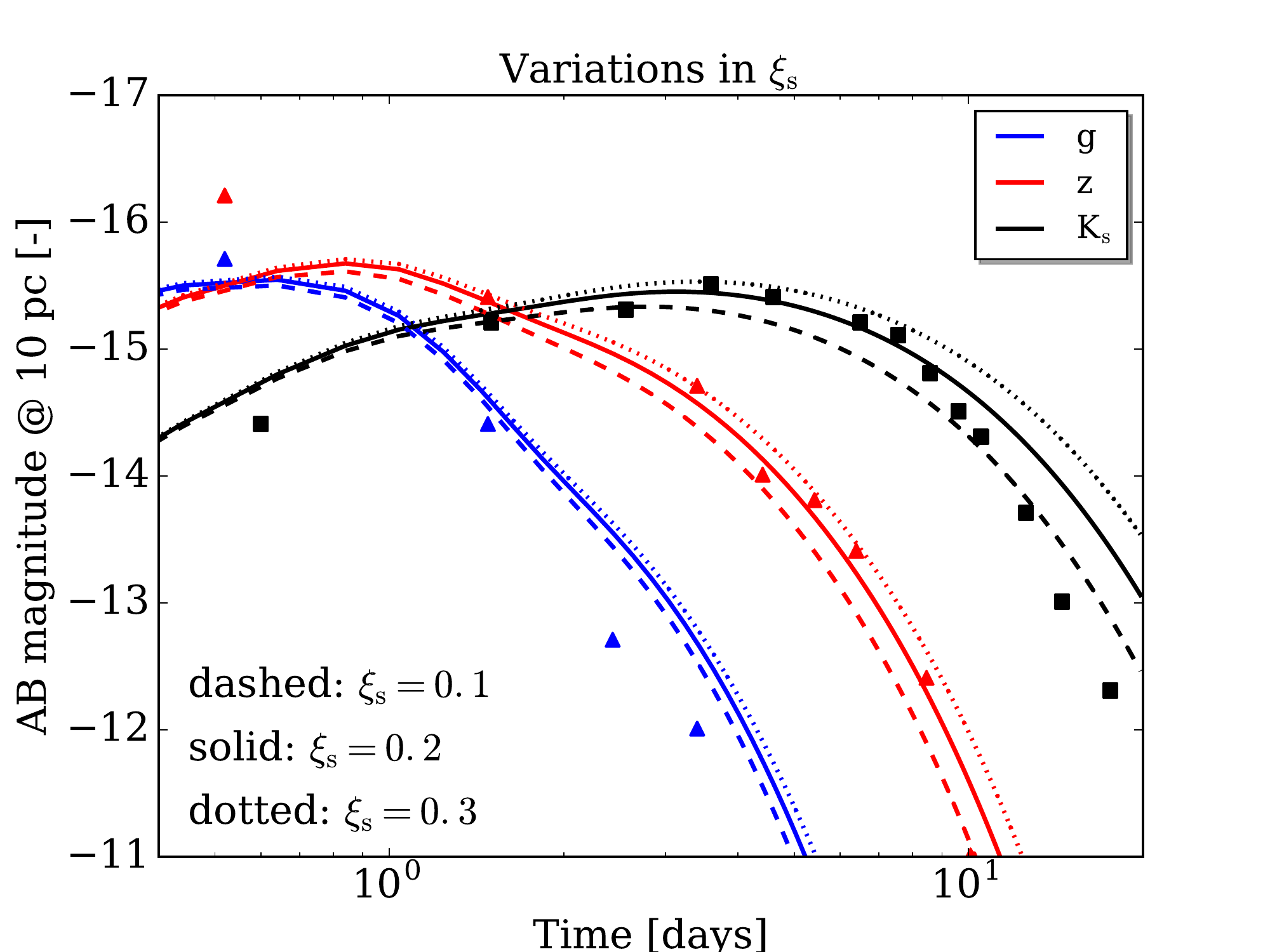} 
 \includegraphics[width=0.32 \linewidth]{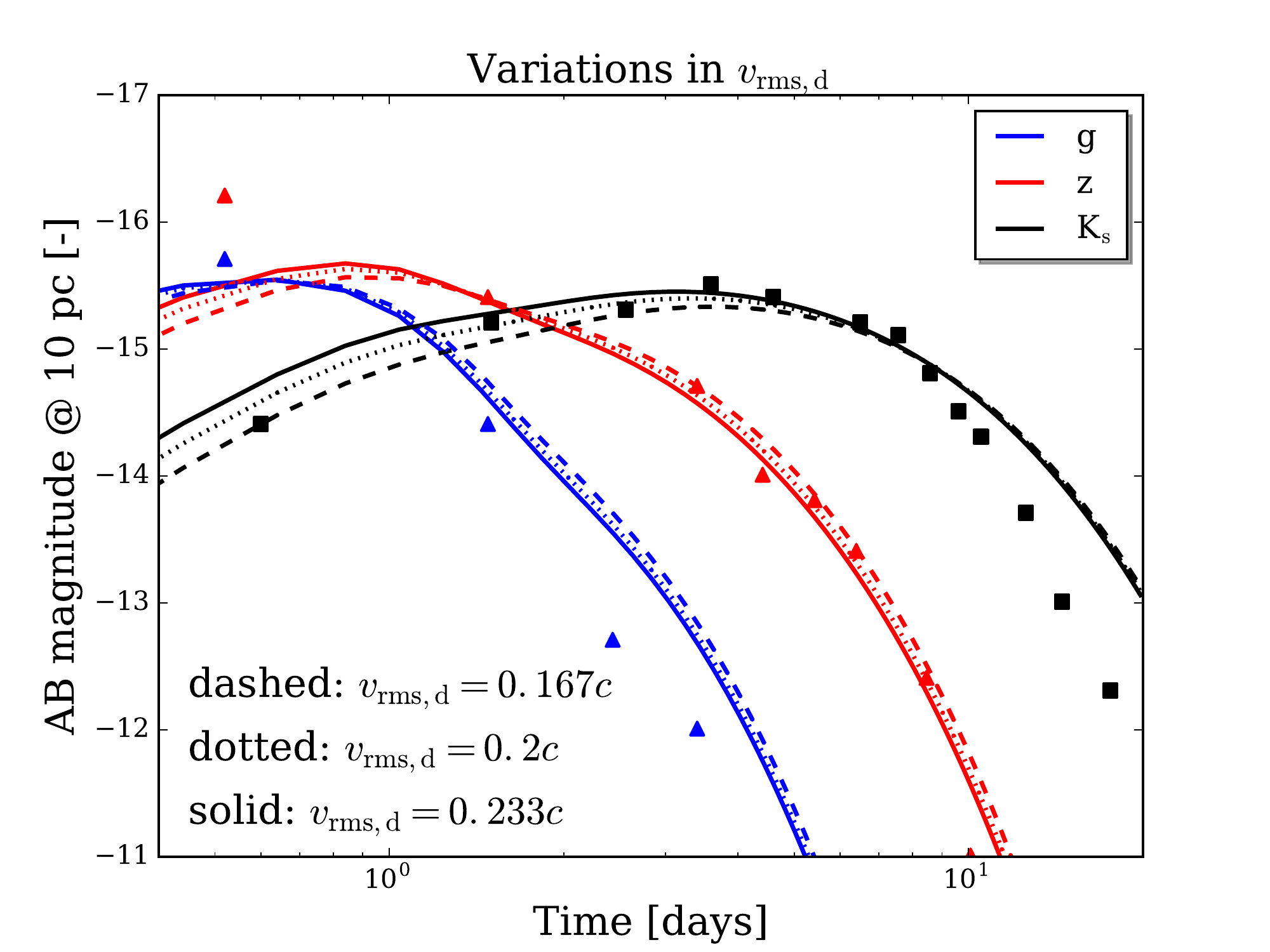}
 \includegraphics[width=0.32 \linewidth]{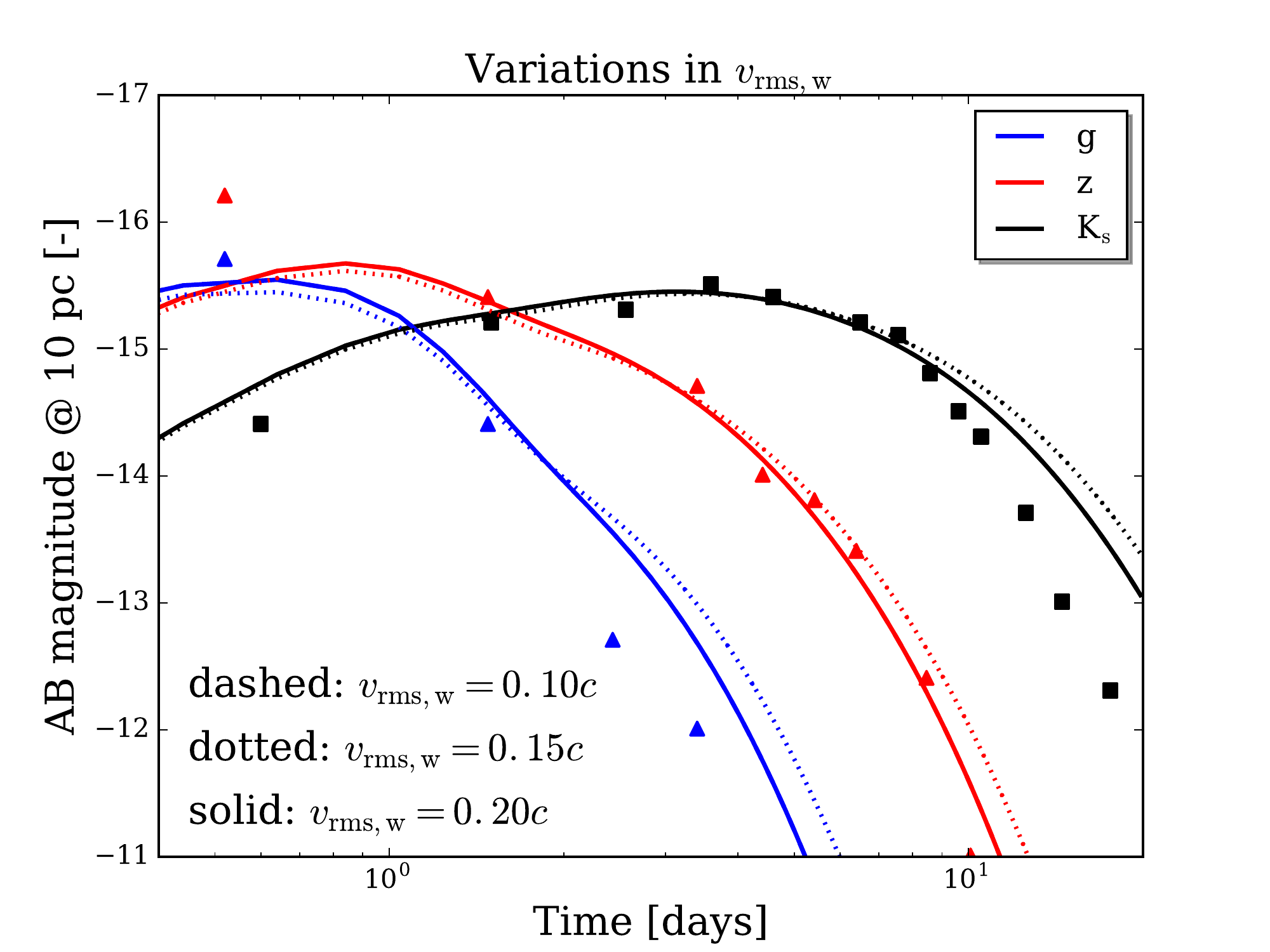}
 \includegraphics[width=0.32 \linewidth]{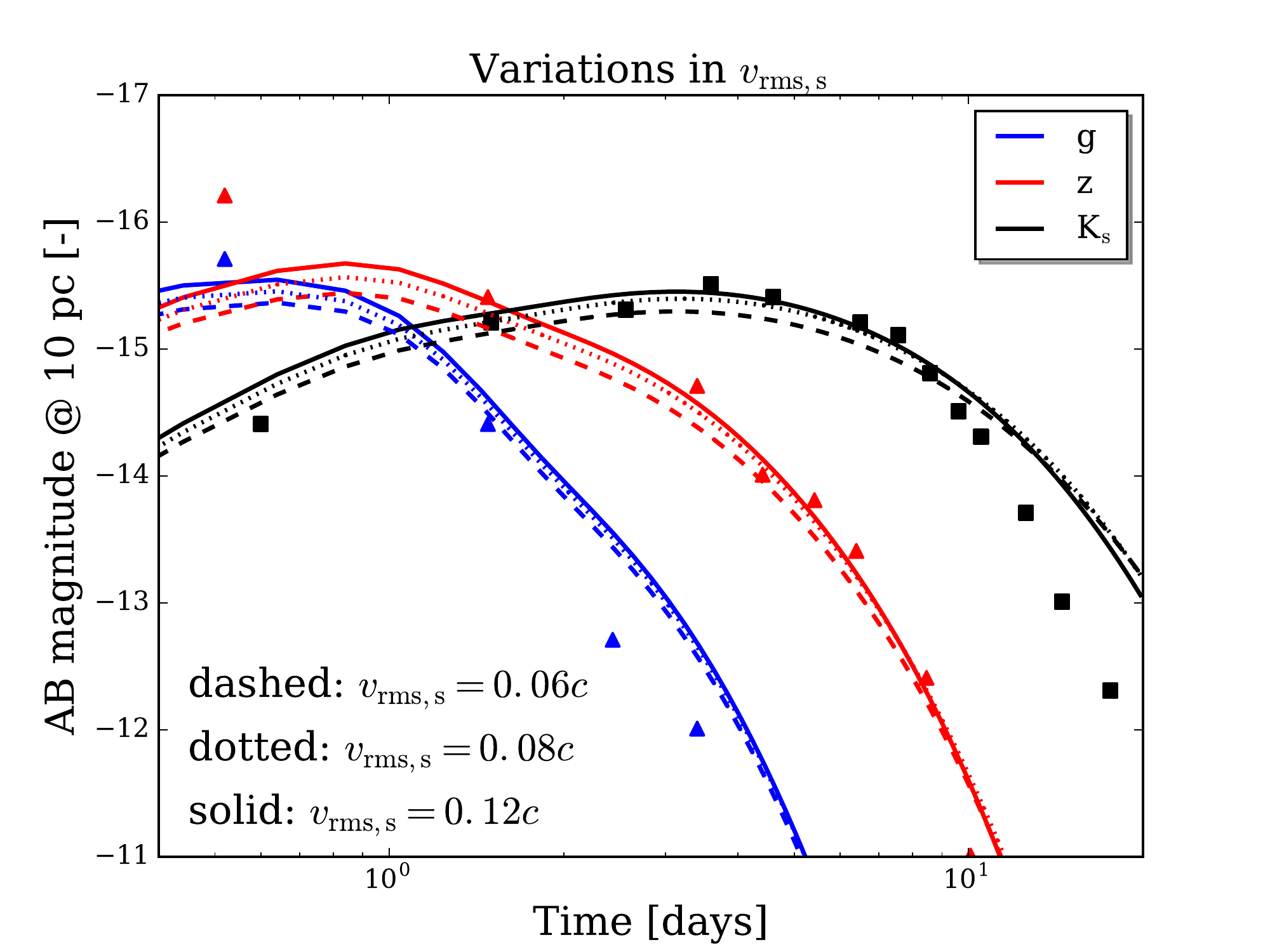}
 \includegraphics[width=0.32 \linewidth]{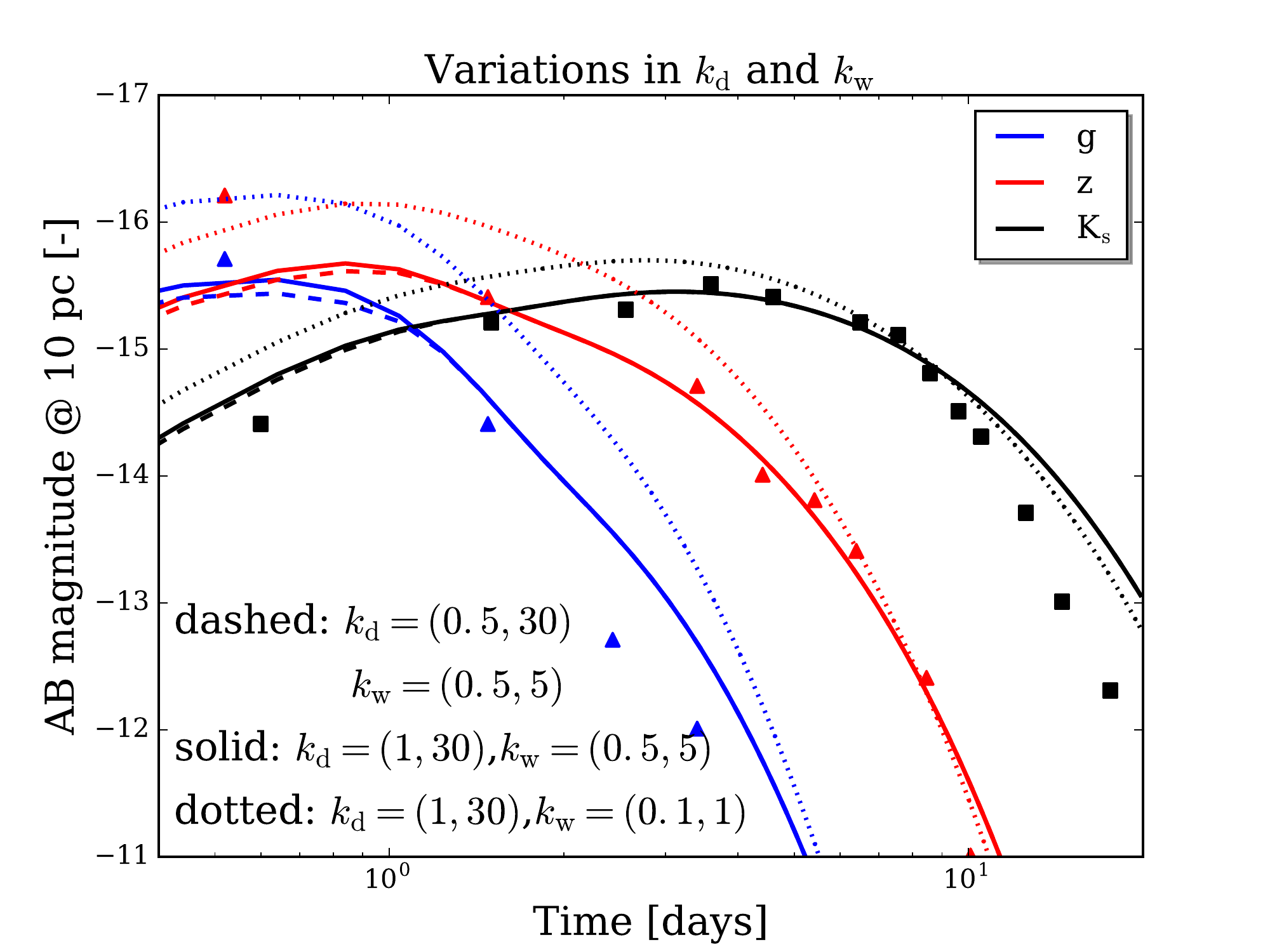}
 \includegraphics[width=0.32 \linewidth]{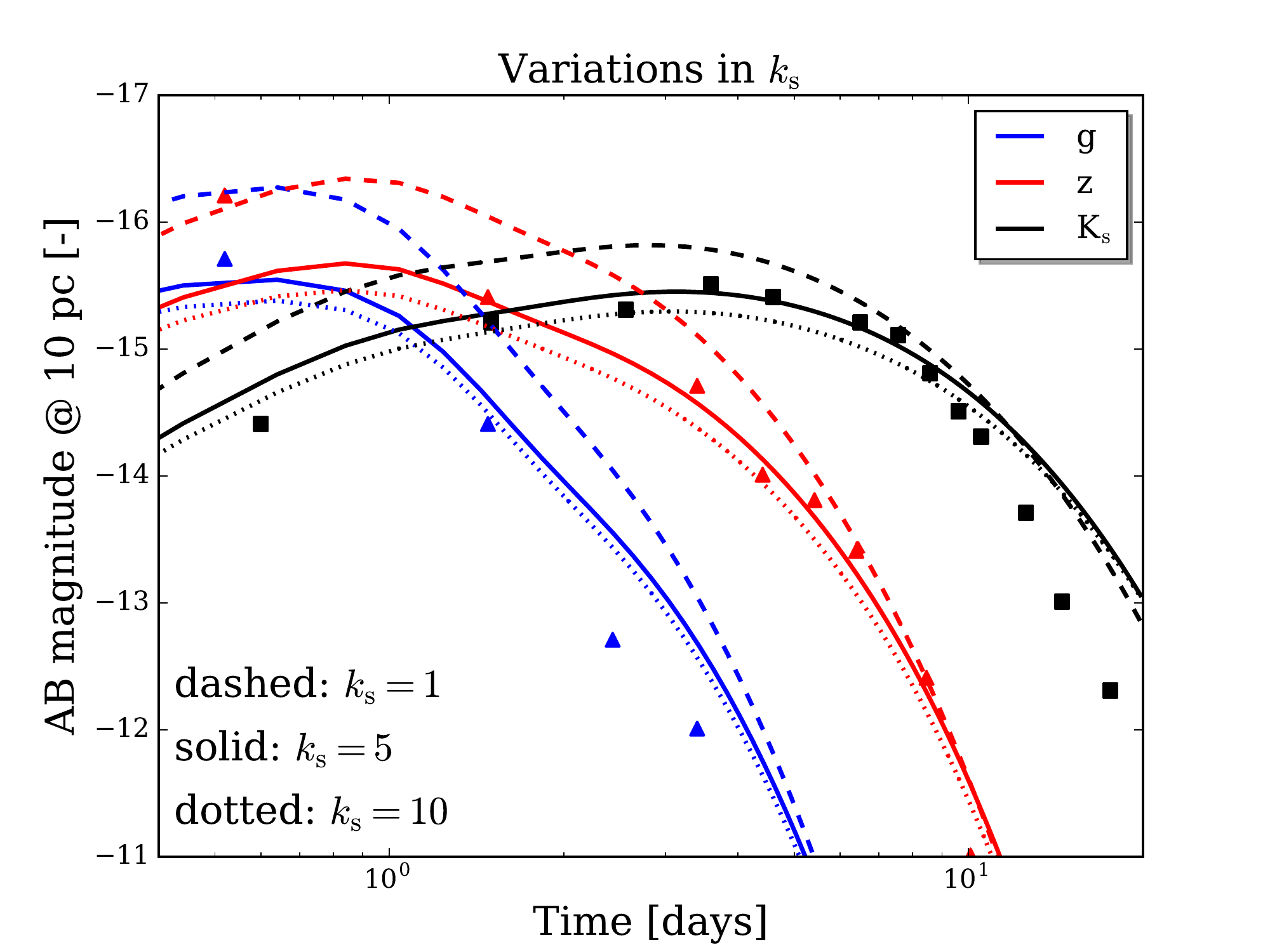}
 \includegraphics[width=0.32 \linewidth]{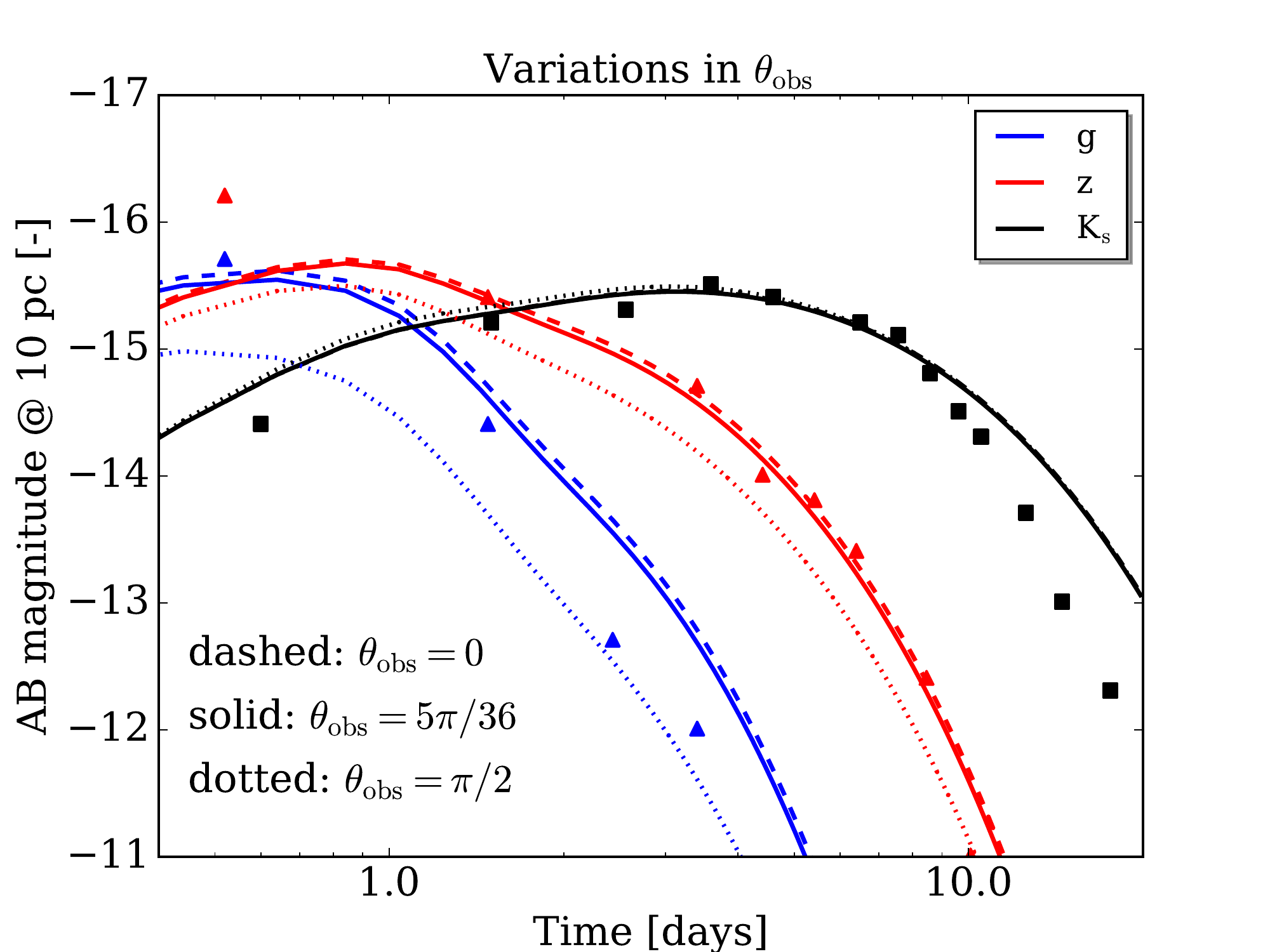}
 \caption{Light curves for the three filters $g$, $z$, $K_{\rm s}$ for several models obtained varying independently one parameter
 (dashed and dotted) with respect to the BF$_{\rm c}$ model (solid).}
\end{figure*}

Due to the peculiar conditions required to underproduce lanthanides and actinides in the secular ejecta (i.e., an extremely long-lived MNS, 
\citealt{Lippuner.etal:2017}), we repeat the analysis by imposing the production 
of a significant amount of heavy r-process elements in this ejecta, i.e. assuming 
$0.1 \lesssim Y_{\rm e,s} \lesssim 0.4$ and $\kappa_{\rm s} \geq 5.0 \, {\rm cm^2 \, g^{-1}}$. 
This constraint alone would result in models with $M_{\rm disk} \geq 0.2 \msun$ and $m_{\rm ej,d} \geq 2 \times 10^{-2} \msun$,
in tension with results from GRHD simulations.
Hence, we impose two more constraints: $m_{\rm ej,dyn} \leq 0.01~\msun$ 
and $M_{\rm disk} \leq 0.12~\msun$.
Under these additional hypothesis, we obtain a new parameter set (BF$_{\rm c}$ in Table~\ref{tab: best fit parameters}) 
whose fit quality has decreased compared with BF ($\chi^2 = 1263$).
% , corresponding to an average discrepancy of 0.124 magnitudes per point),
% % \bs{if you give the avg value here you should give it also for BF}) \alp{I gave it above},
% but still comparable with observational uncertainties. 
The dashed lines in Figure~\ref{fig3} show the corresponding light curves.
In this model, the wind component is now present and well distinct from the 
secular ejecta. Moreover, the opacity for the polar dynamic and polar wind ejecta are comparable.
The amount of wind ejecta (relative to the disk mass) obtained for BF$_{\rm c}$ is significantly above the results 
reported in \cite{Martin.etal:2015} for pure neutrino-driven winds, suggesting a non-negligible role of magnetically-driven winds,
while the amount of secular ejecta is in agreement with the results reported in 
\cite{Just.etal:2015a}, \cite{Fernandez.Metzger:2013} and \cite{Siegel.Metzger:2017}.
In Figure~\ref{fig1}, we explore the sensitivity of our model by varying independently a single parameters 
with respect to the BF$_{\rm c}$ set.
The different panels show that the most relevant light curve features (e.g., peak strength and time, decline behavior) are primarily 
influenced by the total amount of emitting matter and by the time when matter becomes transparent.

For both BF and BF$_{\rm c}$, the observed brightness requires a heating rate larger 
than the heating rate predicted by theoretically nuclear mass models, 
even within presently nuclear uncertainties (e.g. \citealt{Rosswog.etal:2017}).
To explore this uncertainty, we search for the minimum $\chi^2$ imposing
an additional constraint on the nuclear heating rate, $\epsilon_0 \leq 1.2 \times 10^{19} \, {\rm erg \,g^{-1} \,s^{-1}}$,
still compatible with nuclear mass models \citep{Duflo.Zuker:1995}.
The result is reported in Table~\ref{tab: best fit parameters} as BF$_{\rm c,\epsilon}$ and in Figure~\ref{fig3} as dotted lines. 
The agreement with the observations further reduces and $\chi^2$ increases by $\sim$15\%.
Most of the model parameters remain the same as for BF$_{\rm c}$, while
the reduced heating rate is compensated by an increase in the fraction of the disk 
ejected as wind or secular ejecta.

For all best-fit \MKN{} models, the emission is produced by a substantial amount of ejecta: 
$m_{\rm ej} \equiv (m_{\rm ej,d} + m_{\rm ej,w} + m_{\rm ej,s}) = 0.0421$, $0.04~\msun$, and
$0.077 \msun$ for the BF, BF$_{\rm c}$ and BF$_{\rm c,\epsilon}$, respectively.
Our models favor a viewing angle $ \pi/12 \leq \theta_{\rm obs} \leq 7 \pi /36 $, 
with the lower bound (more consistent with GW170917) 
characterized by the presence of a smaller amount of mass ejected along the polar direction.
% and BF$_{\rm c,\epsilon}$ cases).. (here we have to change something...)}
% with the lower bound (more consistent with GW170917) 
% related with the presence of the wind close to the polar axis (BF$_{\rm c}$ 
% and BF$_{\rm c,\epsilon}$ cases).}
Variations of $\theta_{\rm lim,d(w)}$ between $\pi/6$ and $\pi/4$ have a minor impact on our results, but more 
collimated wind outflows are more compatible with smaller $\theta_{\rm obs}$. 
Finally, the presence of a larger nuclear heating rate for the high-$Y_{\rm e}$, polar ejecta
at $t \lesssim t_{\epsilon}$ increases the light curves by half a magnitude during the first
day. Thus, this correction is potentially relevant to explain the early behavior of the UV and visible 
light curves of a \MKN{}.

\section{Conclusion}

% one sentence to say what we did
In this \textit{Letter}, we have interpreted
\AT{}, the EM counterpart of GW170817, 
as the \MKN{} emission produced by a multi-component and 
anisotropic distribution of the ejecta from a BNS merger.

The emission brightness requires a high nuclear heating rate
in combination with an ejected mass in excess of 0.04~$\msun$.
A heating rate compatible with present nuclear uncertainties
implies an even larger mass ejection, 0.077~$\msun$.
The amount of dynamic ejecta predicted by our models ($\sim 0.005 - 0.01~\msun$) is 
consistent (within present uncertainties) with
% systematically larger than the 
typical values provided by GRHD simulations.
% but still compatible with uncertainties in the simulations.
Secular and wind ejecta play a central role and demand the presence of a disk 
with $M_{\rm disk} \gtrsim 0.08~\msun$. The formation of such disks, compatible with 
numerical results, excludes that the merger outcome is a prompt collapse to a BH.

The presence of a BC in the \MKN{} light curve 
is a signature of fast expanding, low opacity ejecta close to the polar region. 
However, reproducing its properties in combination 
with the ones of the RC requires the presence of matter with an 
opacity lower than 10~${\rm cm^2 \,g^{-1}}$ and a more isotropic distribution, in addition 
to very opaque ejecta expected from the equatorial dynamic ejecta.
These results indicate that weak processes are key to set the properties of a fraction of 
the ejecta and have a direct impact on the EM counterpart of BNS mergers.

The ratio between the magnitudes of the BC and RC also constrains %in our model also 
the observer viewing angle to be $\pi/12 \lesssim \theta_{\rm obs} \lesssim 7 \pi/36$.
This interval is fully consistent with the broader limit inferred from the 
GW signal alone ($ \theta_{\rm obs} \lesssim 11 \pi/36$), while the limit including the 
information about the host galaxy distance (NGC4993, $ \theta_{\rm obs} \lesssim 7 \pi/45$)
is more consistent with an observer location at $ \pi/12 \lesssim \theta_{\rm obs} \lesssim 5 \pi /36$.

The observed light curve of \AT{} is compatible both with a (effective) two- and a
three-component ejecta model.
The presence of a very long-lived MNS, necessary to explain the $Y_{\rm e}$ distribution 
required by the two-component model, implies a very efficient mechanism to prevent angular 
momentum redistribution and the subsequent collapse of the MNS (in particular for soft nuclear EOS). 
The three-component models require a reduced amount of dynamic ejecta, in association with 
larger viewing angles (but still compatible with GW170817 constraints). The presence of 
a massive disk and of a wind component still demands a MNS phase, but for a timescale 
shorter than the disk lifetime. 

In our analysis, we did not consider the possible presence of a small amount ($\sim 10^{-4}~\msun$) 
of free-neutron ejecta \citep{Bauswein.etal:2013}, which could contribute to the UV/visible emission 
a few hours after the merger \citep{Metzger.etal:2015}.
We anticipate that the inferred amount of polar outflow would decrease
in the presence of this neutron skin, and we postpone its
detailed study to a future work.

According to our models, the disk ejecta expands significantly slower
than the dynamic ejecta. Its nucleosynthesis yields have more chances not to 
escape from the galaxy and to contribute to its metal enrichment. This 
could help explaining metal abundances in dwarf galaxies \citep{Ji.etal:2016}. 

The discovery of \AT{} has represented a milestone in modern astrophysics. 
Our work indicate that further analysis
of such event will require including the influence of geometry, a
detailed modeling of weak interactions in the merger aftermath (possibly accounting for 
neutrino oscillations, \citealt{Zhu.etal:2016}), and sophisticated GRHD models of BNS mergers that include
viscosity effects and winds \citep{Shibata.Kiuchi:2017,Radice:2017}.

\acknowledgments
The authors thank O. Salafia, T. Venumadhav, and M. Zaldarriaga for useful discussions.
A.P. and D.R. acknowledge support from the 
Institute for Nuclear Theory (17-2b program).
S.B. acknowledges support by the EU H2020 under ERC Starting Grant, 
no.~BinGraSp-714626. 
D.R. acknowledges support from a Frank and Peggy Taplin
Membership at the Institute for Advanced Study and the
Max-Planck/Princeton Center (MPPC) for Plasma Physics (NSF PHY-1523261)
Computations were performed the on supercomputers Marconi/CINECA 
(PRACE proposal 2016153522), Stampede, Comet, and
Bridges (NSF XSEDE allocation TG-PHY160025), on NSF/NCSA Blue Waters
(NSF PRAC ACI-1440083), and PizDaint/CSCS (ID 667).

\bibliographystyle{yahapj}
% \bibliography{bibliography}

\end{document}